\documentclass[12pt,preprint]{aastex}
\shorttitle{Pulsed Gamma-rays from MSPs in 47 Tucanae}
\shortauthors{Venter & de Jager}

\begin{document}
\title{Constraining A General-Relativistic Frame-Dragging\\
Model for Pulsed Radiation from a Population of Millisecond\\
Pulsars in 47~Tucanae using GLAST/LAT}

\author{C. VENTER\altaffilmark{1,2} AND O.C. DE JAGER\altaffilmark{1,2,3}}
\email{Christo.Venter@nwu.ac.za, Okkie.deJager@nwu.ac.za}
\altaffiltext{1}{Unit for Space Physics, North-West University, Potchefstroom Campus, Private Bag X6001, Potchefstroom 2520, South Africa}
\altaffiltext{2}{The Centre for High Performance Computing, CSIR Campus, 15 Lower Hope Street, Rosebank, Cape Town, South Africa}
\altaffiltext{3}{South African Department of Science and Technology, and National Research Foundation Research Chair: Astrophysics and Space Science}

\begin{abstract}
Although only 22~millisecond pulsars (MSPs) are currently known to exist in the globular cluster (GC) \object[NGC 104]{47~Tucanae}, this cluster may harbor $30-60$~MSPs, or even up to $\sim200$. In this Letter, we model the pulsed curvature radiation (CR) gamma-ray flux expected from a population of MSPs in \object[NGC 104]{47~Tucanae}. These MSPs produce gamma-rays in their magnetospheres via accelerated electron primaries which are moving along curved magnetic field lines. A GC like \object[NGC 104]{47~Tucanae} containing a large number of MSPs provides the opportunity to study a randomized set of pulsar geometries. Geometry-averaged spectra make the testing of the underlying pulsar model more reliable, since in this case the relative flux uncertainty is reduced by one order of magnitude relative to the variation expected for individual pulsars (if the number of visible pulsars $N=100$). Our predicted spectra violate the EGRET upper limit at 1~GeV, constraining the product of the number of visible pulsars $N$ and the average integral flux above 1~GeV per pulsar. GLAST/LAT should place even more stringent constraints on this product, and may also limit the maximum average accelerating potential by probing the CR spectral tail. For $N=22-200$, a GLAST/LAT non-detection will lead to the constraints that the average integral flux per pulsar should be lower by factors $0.03-0.003$ than current model predictions.
\end{abstract}


\section{Introduction}
\label{sec:intro}
Millisecond pulsars (MSPs) have been detected in several globular clusters (GCs), including the nearby \object[NGC 104]{47~Tucanae} \citep{C00,F03}. Although only 22~MSPs are currently known to exist in \object[NGC 104]{47~Tucanae} \citep[for a recent review, see][]{L03}, it is expected that this cluster may harbor $30-60$~MSPs \citep{CR05,H05}, or even up to $\sim200$ \citep{IFR05}. 

In this Letter, we model the pulsed curvature radiation (CR) gamma-ray flux expected from a population of MSPs in \object[NGC 104]{47~Tucanae} (\S~\ref{sec:pulsed}), and compare it with the GLAST/LAT sensitivity and EGRET upper limits. Our approach differs from similar recent studies \citep[][hereafter BS07, HUM05, \& CT03]{BS07,HUM05,CT03} in a number of respects. We firstly use a fully 3D General-Relativistic (GR) polar cap (PC) pulsar model \citep[see e.g.][]{VdeJ05}, sampling non-thermal gamma-ray radiation from a range of magnetic field lines above the PC. Next, our study involves a population of MSPs, and we sample randomly from this ensemble. We are therefore able to calculate absolute cumulative fluxes expected from the MSPs in \object[NGC 104]{47~Tucanae}, without needing to use approximate spectra, or scale up single MSP predictions or single particle spectra. We are also specifically interested in the range of pulsed spectra expected due to the uncertain geometries and spread in values of the period $P$ and its derivative $\dot{P}$ of these pulsars. 

X-rays from the MSPs in \object[NGC 104]{47~Tucanae} are believed to be of thermal origin, and are likely produced by a return current heating the pulsars' PCs \citep[][CT03]{B06,HM02}. These X-rays may be Compton upscattered to high-energy gamma-rays. However, this inverse Compton scattering (ICS) component is believed to be dominated by the CR component \citep[see e.g.\ Figure~3 of ][]{BRD00}, and we therefore omitted it from our calculations below.

We lastly make some conclusions regarding the possibility of constraining a model of gamma-ray radiation from GCs (\S~\ref{sec:con}), since an ensemble of MSPs in a GC cluster provides a unique opportunity for specifically constraining the GR frame-dragging pulsar model \citep[e.g.][]{HM98} in a geometry-independent way (\S~\ref{sec:dis}).

\section{Pulsed Gamma-Ray Flux}
\label{sec:pulsed}
The measured values of $\dot{P}$ for GC pulsars are affected by their acceleration in the gravitational potential of the GC, and therefore need to be corrected to obtain intrinsic period derivatives $\dot{P}_{\rm int}$. This was done for \object[NGC 104]{47~Tucanae} by \citet{B06} who derived spin-down luminosities $\dot{E}_{\rm rot}$ assuming a King model. From their Table~4, we selected 13 MSPs and calculated their corresponding $\dot{P}_{\rm int}$ using their periods $P$ as given by \citet{F03}, and the expression $\dot{E}_{\rm rot}=4\pi^2I_{\rm NS}\dot{P}_{\rm int}/P^3$, with $I_{\rm NS}$ the moment of inertia. 

The details of the implementation of our isolated MSP model, using the GR-framework of Harding, Muslimov, \& Tsygan \citep[e.g.][]{HM98} may be found in \citet{VdeJ05}, where we discuss the essential link between pulsar visibility and the assumed geometry, i.e.\ magnetic inclination angle $\chi$ and observer angle $\zeta$ (both measured with respect to the spin axis $\mathbf{\Omega}$). 

HUM05 found that most MSPs are inefficiently screened by CR and ICS pairs. Screening caused by electron-positron pairs will lower the accelerating potential, leading to a lower pulsed CR spectral cut-off. Higher numbers of energetic electrons may however boost the unpulsed gamma-ray spectrum. The vast majority of MSPs in our population lie below the critical spin-down luminosity above which screening is expected to occur \citep{HMZ02}
\begin{equation}
\dot{E}_{\rm rot,break}\approx1.4\times10^{34}\left(\frac{P}{B_{12}^2}\right)^{1/7}\quad{\rm ergs/s}.
\end{equation}
Here, $B_{12}=B/10^{12}$~G, and $P$ is the pulsar period in seconds. (Note the positive sign of the power of $P$). This condition may be rewritten so that screening occurs when
\begin{equation}
\log_{10}\dot{P}>2.625\log_{10} P-12.934,
\end{equation}
assuming $I_{\rm NS}=0.4MR^2$, stellar radius $R=10^6$~cm, and pulsar mass $M=1.4M_\odot$.
We used an unscreened electrical potential for the bulk of the population, and used the approximation of a screened electric field given by \citet{DR00} for MSPs with spin-down powers greater than this critical spin-down power (which was only the case for 47~Tuc~U)). 

For each pulsar $i$ in our population, we calculated phase-averaged CR spectra via \citep[see also Eq.~12 of][]{VdeJ05} 
\begin{eqnarray}
\frac{dN^i_\gamma}{dE}(E_\gamma,\chi,\zeta) & = & \frac{1}{2\pi d^2\sin\zeta d\zeta dE_\gamma}\times\\ \, & \, & \int\int_\zeta^{\zeta+d\zeta}\int_0^{2\pi}\frac{I(E_\gamma,E_\gamma+dE_\gamma)}{E^c_\gamma}\left[\frac{dL^i_\gamma(\phi_L,\chi,\zeta,E_\gamma)}{d\phi_L d\zeta dE_\gamma}\right]d\phi_L d\zeta dE_\gamma,
\end{eqnarray}
where $\phi_L$ is the phase angle, and $I(E_\gamma,E_\gamma+dE_\gamma)$ picks out the gamma-ray energies in the range $(E_\gamma,E_\gamma+dE_\gamma)$. Primary electrons leave a stellar surface patch $dS$ relativistically ($\beta_0=v_0/c\sim1$) at a rate of $d\dot{N}_e=\rho_ed S\beta_0 c/e$, where $\rho_e$ is the charge density. These primaries radiate CR at a position $(r,\theta,\phi)$ above the PC with a characteristic energy of $E^c_\gamma=1.5\lambda_c\gamma^3m_ec^2\rho_c^{-1}$, where $\lambda_c=\hbar/(m_ec)$ is the Compton wavelength, and $\rho_c(r,\theta,\phi)$ is the curvature radius of the GR-corrected dipolar magnetic field. The CR is associated with an incremental gamma-ray luminosity of $dL_\gamma = d\dot{N}_e\dot{E}_{\rm CR}dt$, which is radiated in a time $dt$. The CR loss rate is given by $\dot{E}_{\rm CR}\approx2e^2c\gamma^4/(3\rho_c^2)$, where $\gamma(r,\theta,\phi)$ is the Lorentz factor of the electron primaries. We calculated CR photon spectra $dN^i_\gamma/dE(E_\gamma,\chi,\zeta)$ for $\chi=\zeta=10^\circ,20^\circ,\cdots,80^\circ$, and $i=1,2,...,13$, assuming $R=10^6$~cm, $M=1.4M_\odot$, and $I_{\rm NS}=0.4MR^2$. All spectra ($13\times8\times8=832$) were scaled to a distance of $d=5$~kpc \citep{G03}. 

\begin{figure}
\epsscale{0.8}
\plotone{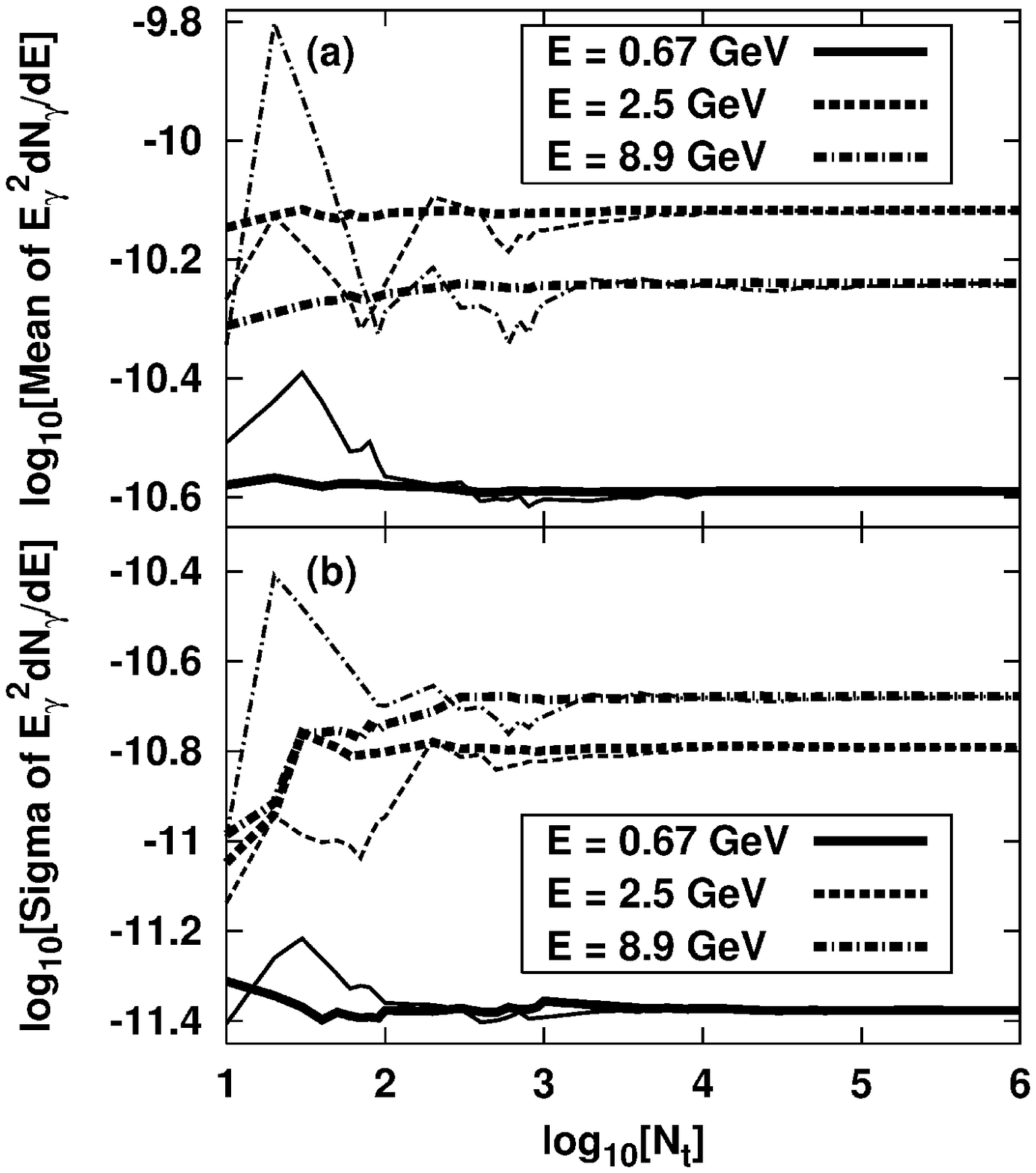}
\caption{The mean and standard deviation $\sigma$ of $E_\gamma^2dN_\gamma/dE$ (in units of ergs/s/cm$^2$) vs.\ $N_t$ as shown in panel~(a) and panel~(b), with $N_t$ the number of values of $E_\gamma^2dN_\gamma/dE$ used to calculate the mean and $\sigma$-values. Line types correspond to different energies as shown in the legend. Thick lines are for values of $E_\gamma^2dN_\gamma/dE$ summed over $N=100$ randomly chosen visible pulsars, while thin lines are for values of $E_\gamma^2dN_\gamma/dE$ for single ($N=1$) randomly chosen pulsars. The latter values were scaled by a factor 100 in panel~(a), and a factor $\sqrt{100}=10$ in panel~(b). Results converge beyond $N_t\sim10^4$. \label{fig:meansigma}}
\end{figure}

We next randomly chose $N=100$ visible MSPs (with random $\chi$ and $\zeta$), and summed their pulsed spectra to obtain a million cumulative spectra from the Monte Carlo process:
\begin{equation}
\left(\frac{dN_{\rm \gamma}}{dE}\right)^j_{\rm cum} = \sum_{k=1}^{N=100}\frac{dN^k_\gamma}{dE}(E_\gamma,\chi,\zeta),
\end{equation}
where for each index $j=1,2,...,N_t=10^6$, a total of $N=100$ spectra (randomly sampled from the abovementioned 832 spectra) were summed. Therefore, $j$ corresponds to a specific choice of $\chi, \zeta,$ and $k$-values when oversampling from 13 to 100~pulsar spectra. This procedure was repeated for 1~MSP instead of 100, i.e.\ setting $N=1$. Upon comparison of these results, one would expect that the mean values of the single MSP differential spectra would scale with $N$, and the standard deviations with $\sqrt{N}$. In panel~(a) of Figure~\ref{fig:meansigma}, the mean values of $E_\gamma^2dN_\gamma/dE$ at three different photon energies $E_\gamma$ are shown as a function of the number $N_t$ of values used to calculate the mean (with a maximum of $N_t=10^6$). The thick lines indicate results when $N=100$, while the thin lines indicate results for $N=1,$ scaled by a factor 100. Similarly, panel~(b) of Figure~\ref{fig:meansigma} indicates the standard deviation of $E_\gamma^2dN_\gamma/dE$ at three different photon energies vs.\ $N_t$. The thick lines again indicate results when $N=100$, while the thin lines indicate results for $N=1$ scaled by a factor 10. It is clear that the mean and $\sigma$-values converge beyond $N_t\sim10^4$, for both $N=1$ and $N=100$. The $N=1$ results are more erratic at small $N_t$-values. The relative error on the mean values are therefore a factor 10 lower when using a population of 100 MSPs, as opposed to the much larger error obtained for single pulsars. Figure~\ref{fig:nuFnu} depicts average cumulative pulsed differential spectra $\left<E^2_\gamma dN_\gamma/dE\right>_{\rm cum}$ for $N=100$ and $N_t=10^6$, as well as $2\sigma$-bands. Also shown are the GLAST/LAT sensitivity curve (HUM05) and EGRET upper limits \citep{F95}, as well as other predictions of pulsed gamma-ray radiation from \object[NGC 104]{47~Tucanae} (HUM05).

\section{Discussion}
\label{sec:dis}
The gamma-ray visibility of an isolated MSP depends extremely sensitively on the assumed pulsar geometry. Inclination and observer angles $\chi$ and $\zeta$ are usually estimated from radio polarization measurements (e.g. \citet{MJ95}), involving the rotating vector model. In our MSP model, the electric potential boundary condition of $\Phi(\xi=1)=0$ is used \citep{MH97} as the last open magnetic field lines $(\xi=1)$ are treated as equipotentials. In addition, the space-charge-limited nature of this model requires that the accelerating electric field parallel to the magnetic field ($E_{||}$) should be zero at the stellar surface $(r=R)$. This means that on-beam radiation (when $\chi\sim\zeta$) will have a spectral cut-off which roughly scales with the maximum potential $\Phi_{\rm max}$. Off-beam radiation will however rapidly decrease as the impact angle $\beta=\zeta-\chi$ increases, because this radiation is produced along field lines where the electric potential $\Phi$, and hence $E_{||}$, has dropped significantly. 

An ensemble of MSPs provides the unique opportunity to side-step the issue of pulsar geometry (which is crucial when modeling a single MSP) by averaging over numerous spectra and geometries, which results in a ``geometry-averaged'' spectrum with relatively small uncertainty. We also exploit this scenario in the sense that we use results from the corotating frames of the MSPs, since we expect that the effect of Lorentz transformations to the observer frame will average out over the ensemble. \textit{We find that the relative spread of ensemble values for the bolometric gamma-ray and electron luminosities as well as the spin-down power and average gamma-ray spectra are typically one order smaller than for the same quantities averaged as single MSP quantities (i.e.\ they scale with $N^{-1/2}$)}. This illustrates the importance of testing pulsar models by observing GCs. (This conclusion follows when using a constant equation of state (EOS). Variations in the EOS have a much smaller effect than that of differing pulsar geometries, and is therefore not included in this discussion; see \S4).

\begin{figure}
\epsscale{0.8}
\plotone{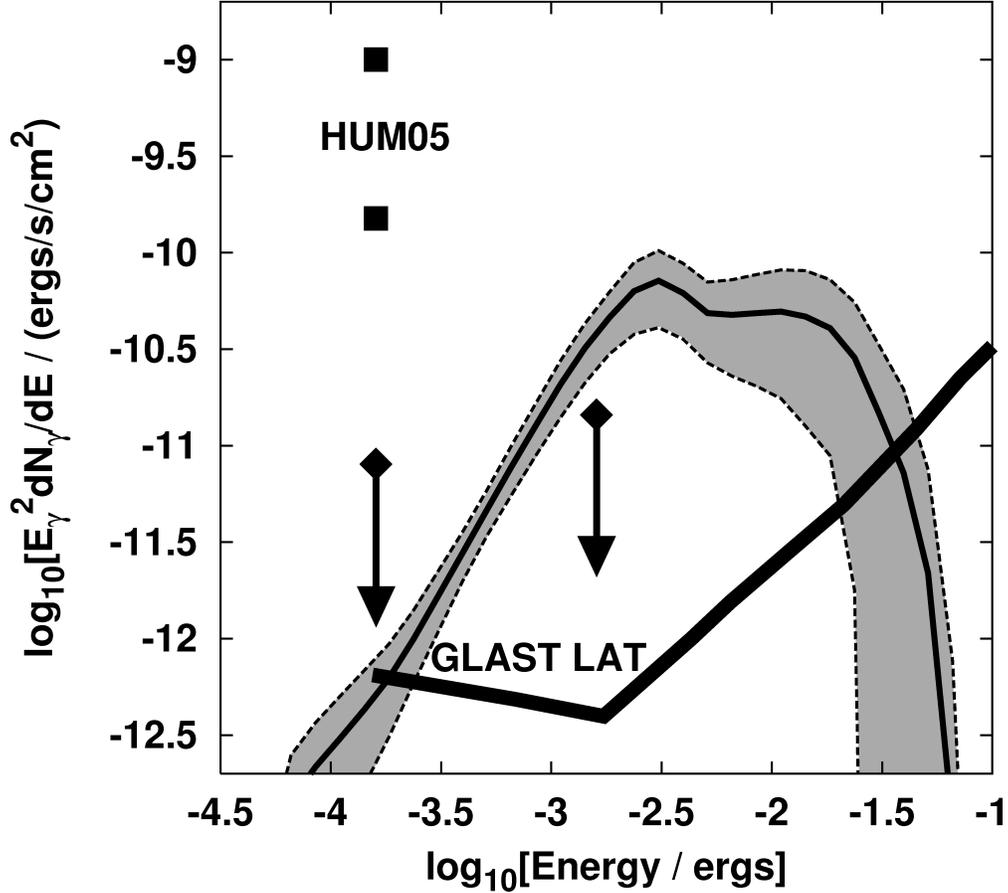}
\caption{Average differential pulsed gamma-ray spectrum (solid line) with $2\sigma$-error bands (shaded band between dashed lines) for $N=100$ and $N_t=10^6$, with smoothed high-energy tails. Furthermore we indicate predictions of HUM05 (lower square for 15~MSPs, upper square scaled to 100~MSPs). The predictions from HUM05 involves estimates of single-particle spectra scaled using the Goldreich-Julian PC current \citep{GJ69} and arbitrary beaming angles. Our spectra include variations across the PC (sampling different magnetic field lines) as well as variations due to various inclination angles and observer geometries (with beaming angles calculated implicitly). The GLAST/LAT sensitivity (HUM05) as well as EGRET upper limits \citep[diamonds -][]{F95} are also shown (the EGRET upper limits and HUM05 results were converted to differential values assuming an $E^{-2}$-spectrum).\label{fig:nuFnu}}
\end{figure}

Due to the lower relative ensemble errors, one should therefore expect to be able to constrain average pulsar parameters using average spectra and their errors shown in Figure~\ref{fig:nuFnu}. This would not be the case if ensemble values did not converge (Figure~\ref{fig:meansigma}) and had relatively small errors. Furthermore, the fact that the ratios of average cumulative luminosities and spin-down powers to their single MSP counterparts are very close to $N=100$ gives us confidence in our sampling algorithm. 

In our model, the predicted average single pulsar efficiencies for electron and gamma-ray production are typically $\sim2\%$ and $\sim7\%$, depending on the population of MSPs used. For our \object[NGC 104]{47~Tucanae} MSP population (with $\left<\dot{E}_{\rm rot}\right>=2.2\times10^{34}$~ergs/s), we found 0.74\% and 6.1\% respectively. This should be compared to the estimate of BS07 of 1\% for the first value, and to that of \citet{HMZ02} of $\sim10\%$ for the latter quantity. \citet{VdeJ05} quoted values of $1-2.5\%$ and $2-9\%$ for these quantities for the case of the pulsar PSR~J0437-4715.

In Figure~\ref{fig:nuFnu}, we converted integral EGRET upper limits \citep{F95} and model predictions of HUM05 to differential values assuming a $E^{-2}$ spectrum. Our predicted pulsed gamma-ray flux at 100~MeV is below the differential EGRET upper limit, much lower than that of HUM05. Their prediction of the integral flux above 100~MeV exceeds the EGRET upper limit by a factor $f\sim19$ for 15 MSPs, and $f\sim125$ for 100 MSPs. (This seemingly overprediction of integral flux was also found in the case of PSR~J0437-4715 \citep{VdeJ05}. HUM05's prediction is a factor $f\sim35$ above the EGRET upper limit for this MSP). However, our summed pulsed spectra violate the EGRET upper limit at 1~GeV \citep{F95}, providing constraints on the low-energy tail of the average pulsed spectrum. GLAST/LAT observations will however constrain the predicted spectrum at all energies of relevance, which should provide a meaningful constraint on the product of visible pulsars $N$, and the average intergral flux above 1~GeV per pulsar. 

\section{Conclusions}
\label{sec:con}
In this Letter, we modeled the pulsed gamma-ray flux expected from a population of MSPs in \object[NGC 104]{47~Tucanae}. We chose a number of $N=100$ visible members, but the spectra can easily be (linearly) scaled to any other reasonable number. We were especially interested in obtaining errors on the pulsed gamma-ray spectrum which would reflect the uncertainty in the geometries, as well as the spread in $P$ and $\dot{P}$, of the individual pulsars. In this way, we wanted to see if it would be possible to constrain the underlying pulsar model independently from the assumed pulsar geometry. It remains to be shown if the spread of the 13 selected pulsars represent the true inherent spread of MSPs in \object[NGC 104]{47~Tucanae}.

For completeness' sake, we furthermore investigated the effect of varying the EOS ($M,R$, and $I$). By choosing $I_{\rm NS}=0.4MR^2$, and letting $M$ and $R$ vary between $M/M_\odot=1.4, 1.5, 1.6$ and $R_6=R/10^6\,{\rm cm}=1.0, 1.2, 1.4, 1.6$, we found that the maximum integral flux did not vary by more than $\sim4\%$ (although it increases with both $M$ and $R$; the spectral cut-off energy furthermore did not vary by more than $\sim13\%$). Since this variation will not significantly influence our constraints derived below, we used fixed values of $M=1.4M_\odot$ and $R_6=1$ throughout.

Our average pulsed spectrum, including errors, exceed the EGRET upper limit at 1~GeV (for the average integral flux spectrum, by a factor of $\sim3.9$). Assuming that this model predicts the gamma-ray flux correctly, this EGRET upper limit constrains the number of GC pulsars to $100/3.9\approx25$. Generally, the constraints derived may be summarised as follows:
\begin{equation}
\frac{N\left<F_{\rm obs}\right>}{\left<F_{\rm model}\right>}\leq \frac{25}{q},
\end{equation}
with $\left<F_{\rm obs}\right>$ and $\left<F_{\rm model}\right>$ the average observed and predicted integral flux above 1~GeV per pulsar, and $q$ is a factor indicating telescope sensitivity. For EGRET, $q=1$, while $q=40$ for GLAST/LAT. E.g., setting $N=22,100,$ and 200, we find that $\left<F_{\rm obs}\right>/\left<F_{\rm model}\right>\leq 1.1, 0.25, 0.13$ for EGRET. This indicates that the EGRET limit is obeyed for $N=22$, but that larger values of $N$ imply a reduction in the model-predicted integral flux (by e.g.\ factors of 4 and 8 for $N=100$ and 200). In the case of a GLAST/LAT non-detection, we find  $\left<F_{\rm obs}\right>/\left<F_{\rm model}\right>\leq 0.03, 0.006, 0.003$ for $N=22,100,$ and 200. GLAST/LAT might lastly even constrain the maximum average electric potential per pulsar $\left<\Phi_{\rm max}\right>$ via the CR cut-off energy. It should also in principle be possible to look for the radio periods corresponding to the 22~known MSPs in \object[NGC 104]{47~Tucanae} in the gamma-ray data (or detect new gamma-ray periods if beaming properties are different), should GLAST/LAT detect the cumulative pulsed CR spectrum.

\acknowledgments
This publication is based upon research supported by the South African National Research Foundation and the SA Centre for High Performance Computing.

\clearpage

\clearpage
\end{document}